\documentclass[12pt]{article}

\newif\iffigs\figstrue

\usepackage{epsfig,latexsym}
\usepackage{amsmath}
\usepackage{color}
\usepackage{verbatim}
\usepackage{mathrsfs}
\usepackage{amssymb}

\DeclareMathAlphabet{\mathpzc}{OT1}{pzc}{m}{it}

 \csname
@addtoreset\endcsname{equation}{section}

\def\gz0{\gamma^{0}}

\def\a{\alpha}

\def\e{\epsilon}

\def\beq{\begin{equation}}
\newcommand{\eeq}[1]{\label{#1}\end{equation}}
\def\bea{\begin{eqnarray}}
\newcommand{\eea}[1]{\label{#1}\end{eqnarray}}
\def\ba{\begin{array}}
\def\ea{\end{array}}
\def\bec{\begin{center}}
\def\ec{\end{center}}
\def\ba{\begin{align}}
\def\ena{\end{align}}

\def\12{\frac{1}{2}}

\usepackage{slashed}

\usepackage{float}

\usepackage{booktabs}
\usepackage{caption}

\usepackage[mathscr]{euscript}

\usepackage{color}
\usepackage{graphicx}
\usepackage{epsf}
\usepackage{graphicx,epsfig}
\usepackage{amsmath}

\usepackage{multirow}

\usepackage{amsmath, amsthm, amssymb}

\usepackage{epsfig}
\usepackage{cite}
\usepackage{color,colordvi}

\newcounter{hran}

\makeatletter
\renewcommand\section{\@startsection {section}{1}{\z@}%
                               {-3.5ex \@plus -1ex \@minus -.2ex}%
                               {2.3ex \@plus.2ex}%
                               {\normalfont\large\bfseries}}
\makeatother

\vspace{0.5cm}

\newcommand{\bi}{\begin{itemize}}
\newcommand{\ei}{\end{itemize}}

\begin{document}

\begin{flushright}
CERN-PH-TH/2014-250\\
\end{flushright}

\vspace{15pt}

\begin{center}


{\Large\sc Generalized Born--Infeld Actions \vskip 12pt and Projective Cubic Curves}\\


\vspace{35pt}
{\sc S.~Ferrara${}^{\; a,b,*}$, M.~Porrati${}^{\; c}$, A.~Sagnotti${}^{\; a,**}$, R.~Stora${}^{\; a,d}$ and A.~Yeranyan${}^{\; e,b}$}\\[15pt]

{${}^a$\sl\small Department of Physics, CERN Theory Division\\
CH - 1211 Geneva 23, SWITZERLAND \\ }
\vspace{6pt}

{${}^b$\sl\small INFN - Laboratori Nazionali di Frascati \\
Via Enrico Fermi 40, I-00044 Frascati, ITALY}\vspace{6pt}

{${}^c$\sl\small CCPP, Department of Physics, NYU \\ 4 Washington Pl., New York NY 10003, USA}\vspace{6pt}

{${}^d$\sl\small Laboratoire d'Annecy-le-Vieux de Physique Th\'eorique (LAPTH) \\
F-74941 Annecy-le-Vieux Cedex, FRANCE}\vspace{6pt}

{${}^e$\sl\small Centro Studi e Ricerche Enrico Fermi\\
Via Panisperna 89A, 00184, Roma, Italy}

\vspace{8pt}

\vspace{35pt} {\sc\large Abstract}
\end{center}
\baselineskip=20pt
We investigate $U(1)^{\,n}$ supersymmetric Born--Infeld Lagrangians with a second non--linearly realized supersymmetry. The resulting non--linear structure is more complex than the square root present in the standard Born--Infeld action, and nonetheless the quadratic constraints determining these models can be solved exactly in all cases containing three vector multiplets. The corresponding models are classified by cubic holomorphic prepotentials. Their symmetry structures are associated to projective cubic varieties.
\vfill
\line(1,0){250}\\
{\footnotesize {$^*$On leave of absence from Department of Physics and Astronomy, U.C.L.A., Los Angeles CA
USA}}\\
{\footnotesize {$^{**}$On leave of absence from Scuola Normale Superiore and INFN, Piazza dei Cavalieri 7, 56126 Pisa ITALY}}

\noindent

\setcounter{page}{1}

\pagebreak

\newpage
\section{Introduction}\label{sec:intro}

Supersymmetric off--shell generalizations of the $N=1$ Born--Infeld (BI) Lagrangian
\cite{BI,deser,CF} have been proposed in the literature for the case of $n$ Maxwell fields. They all share the property of reducing to the standard BI for $n=1$. The latter, however, admits a second non--linearly realized supersymmetry \cite{BG}, and can be interpreted as a low--energy effective action for the partial $N=2 \to N=1$ breaking of rigid supersymmetry. Ref.~\cite{FPS} recently presented a multi--field generalization of the BI action that, unlike the previous proposals in \cite{TSE,RT,BMZ,KT}, also admits a second non--linearly realized supersymmetry \cite{HLP,FGP}. This property makes the construction unique. The goal of the present investigation is to perform a detailed analysis of the $n=3$ case.

The action in \cite{FPS} is built starting from $n$ $N=1$ chiral multiplets and $n$ $N=1$ vector multiplets, which are building blocks of $N=2$ vector multiplets. Our starting point extends the setting that Antoniadis, Partouche and Taylor connected in \cite{APT} to the partial breaking of global $N=2$ supersymmetry. For $n=1$ the construction reduces to the standard $N=1$ BI action. However, for $n>1$ the generalization is non trivial, rests on the holomorphic cubic prepotential of $N=2$ rigid special geometry and also affords a natural interpretation in terms of projective cubic $(n-2)$--varieties.

This article is organized as follows. In Section \ref{sec:BI_actions} we review the non--linear $N=2$ constraints that define the $N=2$ multi--field BI actions. In all $n=2$ \cite{FPS} and, as we shall see, also in the $n=3$ cases, these constraints can be solved explicitly, in spite of the fact that they are coupled systems of $n$ quadratic equations. In Section \ref{sec:orbits} we summarize the classification of projective cubic curves \cite{inv_theory}, whose singularity structure (see Table \ref{tab:clasinv}) underlies the possible independent sets of $N=2$ constraints (see eqs.~\eqref{5} and \eqref{6}). These are connected to the orbits of the three--fold symmetric projective representation $P(Sym^3(R^n))$ of $SL(n,R)$, since the BI constraints are also invariant under an overall rescaling. In particular, for $n=3$ the orbits are classified, up to a rescaling, by the two invariants $P_4$ (of degree four) and $Q_6$ (of degree six) \cite{BS} of $P(Sym^3(R^{\,n}))$ when they do not vanish. This classification proceeds in two steps: first the classification over the complex numbers (Section \ref{sec:complex}), and then its refinement over the real numbers  (Section \ref{sec:real}). Most cases are associated to degenerations of cubic curves where the $SL(3,R)$ invariants $P_4$ and $Q_6$, the discriminant $I_{12} = P_4^{\,3}\,-\,6\,Q_6^{\,2}$, or some of their derivatives vanish. Section \ref{sec:kinetic matrix} is devoted to the contribution of the cubic polynomial to the kinetic matrix, which is generally of the form
\beq
d_{AB} \ =  \ d_{ABC} \ q^C \ ,
\eeq{0}
with real and positive $q^c$, and in particular to its positivity properties. When this Hessian matrix is positive definite, as stressed in \cite{FPS}, a simplification occurs: there is no need to introduce the additional $C_{AB}$ matrix,  and therefore the resulting models are more akin to the standard $n=1$ BI action. The positivity constraints on the Hessian matrix for the different models are summarized in Table \ref{tab:hessian}. The paper ends in Section \ref{sec:conclusion}, which collects our conclusions. Explicit solutions of the BI constraints corresponding to the degenerate families of cubic curves with $(P_4,Q_6) \neq (0,0)$ are collected in Appendix \ref{sec:appendix}, where we illustrate four cases, corresponding to $I_{12}\neq0$, $I_{12}=0$, $\partial I_{12}=0$, and $\partial^{\,2} I_{12}=0$.

\section{$\mathbf{U(1)^n}$ $\mathbf{N=2}$ Born--Infeld actions}\label{sec:BI_actions}

We consider generalized BI Lagrangians that arise from the superspace nilpotency constraints (see ref.\cite{FPS} for more details)
\bea
&& d_{ABC} \left[ W^B\,W^C \ + \ X^B \left( m^C \ - \ \bar{D}^2 \, \label{1} \bar{X}^C\right)\right] \ = \ 0 \ ,  \label{101} \\
&& d_{ABC} \ X^B \, X^C \ = \ 0 \ ,  \\
&& d_{ABC} \ X^B \, W^C \ = \ 0 \ ,
\eea{1001}
and are also invariant under a second, non--linearly realized supersymmetry. This property makes this multi--field generalization very different from the one proposed in \cite{BMZ}, which is not invariant under a second supersymmetry and therefore is not connected to the problem of partial breaking \footnote{In proving the invariance of the action \cite{FPS}, one needs the cubic Fierz identity
\beq
d_{ABC} \ W_\alpha^{\,A} \, W_\beta^{\,B} \, W_\gamma^{\,C} \ = \ 0 \ , \nonumber
\eeq{11}
where $W_\alpha^{\,A} \ = \ \bar{D}^{\,2} \, D_{\alpha} V^{\,A}$
is the chiral field strength of the $N=1$ vector multiplet $V^{\,A}$ \cite{FZ}.}

The bosonic part of the generalized BI actions is determined by the $\theta^{\,2}$--component of the constraints \eqref{1},
\beq
d_{ABC} \left[ G_+^B\cdot G_+^C \ + \ F^B \left( m^C \ - \ \bar{F}^C\right)\right] \ = \ 0 \ ,
\eeq{2}
where the $G_+^A$ are the self--dual curvatures of the Maxwell field strengths, and
\beq
\Re \left( G_+^A\cdot G_+^B\right) \ = \  G^A\cdot G^B \ , \qquad \Im \left( G_+^A\cdot G_+^B \right) \ = \ G^A\cdot {\widetilde{G}}^B \ .
\eeq{3}
Here $F^A$ are the auxiliary--field components of the chiral multiplets $X^A$, whose first components will be denoted by $x^A$.

The real parts of eqs.~\eqref{2} are $n$ quadratic equations that are generally coupled. Letting
\beq
H^A \ = \ \frac{m^A}{2} \ - \ \Re F^A \ , \qquad R^{AB} \ = \ G^A\cdot G^B \ + \frac{m^A\,m^B}{4} \ - \ \Im F^B \, \Im F^C \ ,
\eeq{4}
they take the form
\beq
d_{ABC} \left( H^B\, H^C \ - \ R^{BC} \right) \ = \ 0 \ .
\eeq{5}
On the other hand, the imaginary parts of eqs.~\eqref{2} are $n$ linear equations for $\Im F^A$:
\beq
d_{ABC} \left( G^B \cdot {\widetilde{G}}^C \ + \ \Im F^B \, m^C \right) \ = \ 0 \ .
\eeq{6}

The bosonic portions of the resulting Lagrangians can be expressed in terms of the $F^A$ and of the real magnetic changes $m^A$, which enter the quadratic system \eqref{1}. They also involve the additional complex charges $e_A=e_{1A}+i\,e_{2A}$, as
\beq
{\cal L}_{\rm Bose} \ = \ e_{2A} \left( \frac{m^A}{2} \ - \ H^A \right) \ + \ C_{AB} \left( H^A \, H^B \ - \ R^{AB} \right) \ + \e_{1A}\, \Im F^A \ .
\eeq{7}

As explained in \cite{FPS}, the real symmetric matrix $C_{AB}$ is needed whenever the matrix $d_{AB}$ in eq.~\eqref{0} is not positive definite. Moreover, by a change of symplectic basis one could also eliminate the real parts $e_{1A}$ of the electric charges, which multiply total derivatives. These Lagrangians combine, in general, a quadratic Maxwell--like term with additional higher--order contributions. For $n=1$, or whenever the matrix $d_{AB}$ of eq.~\eqref{0} is positive definite, one is not compelled to introduce the $C_{AB}$ and the Lagrangian takes the simpler form
\beq
{\cal L}_{\rm Bose} \ = \ e_{2A} \left( \frac{m^A}{2} \ - \ H^A \right) \ + \e_{1A}\, \Im F^A \ ,
\eeq{8}
where the second contribution is a total derivative. In all cases, however, the difficult step in the construction of the Lagrangians is the solution of the quadratic constraints, and in particular of the non--linear ones given in eq.~\eqref{5}.

In principle, for $n=3$ the system should lead to an eight--order equation, which cannot be solved algebraically in general. However, as we shall see, only in the case with $I_{12}\neq 0$ is one led to a quartic equation, while in all other cases with $I_{12}=0$ the system \eqref{5} leads to cubic or biquadratic equations, or even to simple radicals.

Let us stress that the $U(1)^n$ actions of eq.~\eqref{7} are very different from others proposed earlier in the literature. In fact, we claim that they are the only ones that are invariant under a second non--linear supersymmetry, even if only $N=1$ supersymmetry is manifest. A second supersymmetry of the form
\bea
\delta W_\a^{\,A} &=& \left(m^{\,A} \ - \  b \, \ \bar{D}^2 \bar{X}^{\,A} \right) \eta_\a \ - \ 4\, i\, b \ \partial_{\a \bar{\a}}X^{\,A} \ \bar{\eta}^{\bar{\a}} \ , \label{33} \\
\delta X^{\,A} &=& - \ 2 \ W^{\,\a\,A} \ \eta_{\,\a}
\eea{9}
was in fact preserved all the way in our construction, where we started from a linear model with non--renormalizable couplings, an $n$--field generalization of the one proposed in \cite{APT}. Let us mention here that the goldstino superfield is \cite{FPS}
\beq
W_{g\,\alpha} \ = \ \left( e_{2A}\,m^A \right)^{\,-\,\frac{1}{2}} \ e_{2B}\, W_\alpha^B \ ,
\eeq{10}
which indicates that the $N=2$ supersymmetry breaking scale is
\beq
E \ = \ \left( e_{2A}\, m^A \right)^\frac{1}{4} \ .
\eeq{101}
 \section{Invariant polynomials and orbits of the $\underline{10}$ of ${\mathbf SL(3,R)}$} \label{sec:orbits}

The cubic polynomials that define the special geometry are of the form
\beq
U \ = \ \frac{1}{3!} \ d_{ABC} \ x^A\,x^B\,x^C \ ,
\eeq{111}
with $d_{ABC}$ real. For $n=3$ we shall let $x^1=x$, $x^2=y$ and $x^3=z$, and our analysis will rest on the classification of homogeneous cubic polynomials over $R(x,y,x)$ presented in \cite{inv_theory,GG,BS}. We shall encounter 15 distinct types of polynomials over $R$, which were classified according to the different degenerations of the cubic curves defined by $U=0$. The classification rests on the discriminant of the cubic, $I_{12}$, a polynomial of degree 12, and on the two invariants $P_4$ and $Q_6$  \cite{BS},
\begin{eqnarray}
P_4&=&d_{a_1\,a_2\,a_3}\,d_{b_1\,b_2\,b_3}\,d_{c_1\,c_2\,c_3}\,d_{d_1\,d_2\,d_3}\epsilon^{b_1\, a_1\, d_1}\epsilon^{c_2\, d_2\, a_2}
\epsilon^{b_3\, c_3\, a_3}\epsilon^{d_3\, c_1\, b_2}\, ,\\
Q_6&=&d_{a_1\,a_2\,a_3}\,d_{b_1\,b_2\,b_3}\,d_{c_1\,c_2\,c_3}\,d_{d_1\,d_2\,d_3}\,d_{f_1\,f_2\,f_3}\,d_{h_1\,h_2\,h_3}\epsilon^{h_3\, a_1\, b_1}\epsilon^{f_3\, c_1\, a_2}
\epsilon^{d_3\, b_2\, c_2}\epsilon^{c_3\, f_2\, d_2}\epsilon^{a_3\, h_2\, f_1}\epsilon^{b_3\, d_1\, h_1}\, ,\nonumber
\end{eqnarray}
of degrees 4 and 6, built out of the $\underline{10}$ of $SL(3,R)$, which determine $I_{12}$ according to
\beq
I_{12} \ = \ P_4^3 \ - \ 6\, Q_6^2 \ ,
\eeq{12}
in a given normalization convention.

For a generic non--singular cubic $I_{12} \neq 0$, while for singular ones $I_{12}=0$. The singular cases can be further classified in terms of three types of degenerations, depending on whether the curve $U=0$ has singular points (A), is the product of a conic and a line, $U=L \times Q$ (B), or is a product of three lines, $U = L \times L \times L$. As in other contexts, it is convenient to begin by considering the classification over the complex numbers before turning to its finer counterpart over the reals.

 \subsection{Classification over the complex}\label{sec:complex}

Over the complex there are eight degenerate cases \cite{JH}. Two of them are of type (A), that is singular points which are either a node or a cusp. Two are of type (B), and one can distinguish them further according to whether the line intersects the conic or is tangent to it. Finally, four are of type (C), and one can distinguish them further according to whether the lines intersect pairwise (triangle), or all intersect at the same point, or two are concurrent and intersect the third, or finally all three are concurrent.
\begin{figure}[h]
\begin{center}
\epsfig{file=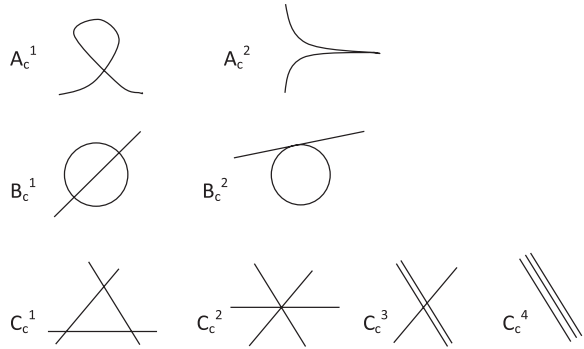, height=1.8in, width=3in}
\end{center}
\caption{\small
Different types of degenerations for complex curves, $I_{12}=0$.
}
\label{fig:sing_curves}
\end{figure}

All these cases fall in two groups, distinguished by the pair of values of $P_4$ and $Q_6$ (see Table \ref{tab:clasinv}) . When these do not vanish, $\partial I_{12}$ and $\partial^{\,2}I_{12}$ may or may not vanish. Moreover, when $P_4$ and $Q_6$ both vanish, different derivatives of these invariant polynomials may or may not vanish.

There is one case with $P_4$ and $Q_6$ not simultaneously zero in each of the (A), (B) and (C) cases of degenerations. In (A) it is the node $(\partial I_{12} \neq 0)$, in $(B)$ it is the line intersecting the conic in two points $(\partial I_{12} = 0)$, while in (C) it is the triangle $(\partial I_{12} \, = \, \partial^{\,2} I_{12}=0)$. There are five cases with $(P_4,Q_6)=(0,0)$, one in (A), one in (B) and three in (C). In (A) there is the cusp, with $\partial P_4$ and $\partial Q_6$ both different from zero, while in (B) there is the line tangent to the conic $(\partial P_4 \neq 0,\partial Q_6=0)$. Finally, in (C) there are three cases: $L \times L \times L$ intersecting at a single point, $(\partial \, P_4 , \partial\, Q_6)=(0,0)$, two concurrent lines (again $(\partial \, P_4 , \partial\, Q_6)=(0,0)$, but $(\partial^{\,2} \, P_4 \neq 0 ,\, \partial^{\,2} \, Q_6=0)$), and three concurrent lines (also $(\partial^{\,2} \, P_4 , \partial^{\,2} \, Q_6) = (0,0)$). Note that all nonzero values of $P_4$ and $Q_6$ in Table \ref{tab:clasinv} correspond to the same value, 6, of the ratio $P_4^3/Q_6^2$, since $I_{12}=0$. For $I_{12} \neq 0$ there is a one--parameter family of projective invariants (see eq.~\eqref{126}).

For each of these models there is a representative normal form, which allows a systematic exploration of different realizations of special geometry  with three vector multiplets. The Fayet--Iliopoulos terms of the $N=2$ theory are in this case $SU(2)$ triplets of $Sp(6,R)$ charge vectors \cite{FPS}.
\begin{table}[tbp]
\begin{center}
\begin{tabular}{|c|c||c||c|c|c|c|c|c|c|c|}
\hline
\rule[-1mm]{0mm}{6mm}$ {R} $ & $ {C}$ & $\rm Polynomial$ & $P_4$ & $Q_6$ & $\partial I_{12}$ & $\partial P_4 $ & $\partial Q_6 $ & $\partial^2 I_{12}$ & $\partial^2 P_{4}$ & $\partial^2 Q_{6}$ \\ \hline\hline
\rule[-1mm]{0mm}{8mm}$A^1$& $A^1_c$ & $-x^3-x^2z+y^2z$ & $\frac{8}{27}$ & $\frac{16}{243}$ & $\neq 0$ & $\neq 0$ & $\neq 0$ & $\neq 0$ & $\neq 0$ &
$\neq 0$ \\
\hline
\rule[-1mm]{0mm}{8mm}$A^2$& $A^1_c$ & $-x^3+x^2z+y^2z $ & $\frac{8}{27}$ & $-\frac{16}{243}$ & $\neq 0$ & $\neq 0$ & $\neq 0$ & $\neq 0$ & $\neq 0$ & $\neq 0$ \\ \hline
\rule[-1mm]{0mm}{8mm}$A^3$& $A^2_c$ &$-x^3+y^2z $ & $ 0$ & $ 0$ & $ 0$ & $\neq 0$ & $\neq 0$ & $\neq 0$& $\neq 0$ & $\neq 0$\\ \hline
\rule[-1mm]{0mm}{8mm}$B^1$& $B^1_c$ &$(x+y+z)(x^2+y^2+z^2) $ & $\frac{8}{3}$ & $-\frac{16}{9}$ & $ 0$ & $\neq 0$ & $ \neq 0$ & $\neq 0$ & $\neq 0$ & $\neq 0$\\ \hline
\rule[-1mm]{0mm}{8mm}$B^2$& $B^1_c$ &$x(x^2+y^2-z^2) $ & $\frac{8}{27}$ & $\frac{16}{243}$ & $ 0$ & $\neq 0$ & $ \neq 0$ & $\neq 0$ & $\neq 0$ & $\neq 0$ \\
\hline
\rule[-1mm]{0mm}{8mm}$B^3$& $B^1_c$ &$z(x^2+y^2-z^2) $ & $\frac{8}{27}$ & $-\frac{16}{243}$ & $ 0$ & $\neq 0$ & $ \neq 0$ & $\neq 0$ & $\neq 0$ & $\neq 0$ \\
\hline
\rule[-1mm]{0mm}{8mm}$B^4$& $B^2_c$ &$(x+z)(x^2+y^2-z^2) $ & $0$ & $0$ & $ 0$ & $\neq 0$ & $0$ & $0$ & $\neq 0$ & $\neq 0$ \\
\hline
\rule[-1mm]{0mm}{8mm}$C^1$& $C^1_c$ &$6 x y z $ & $24$ & $48$ & $ 0$ & $\neq 0$ & $ \neq 0$ & $ 0$ & $\neq 0$ & $\neq 0$ \\
\hline
\rule[-1mm]{0mm}{8mm}$C^2$& $C^1_c$ &$x(y^2+z^2)$ & $\frac{8}{27}$ & $-\frac{16}{243}$ & $ 0$ & $\neq 0$ & $ \neq 0$ & $ 0$ & $\neq 0$ & $\neq 0$ \\
\hline
\rule[-1mm]{0mm}{8mm}$C^3$& $C^2_c$ &$xy(x+y)$ & $ 0$ & $ 0$ & $ 0$ & $ 0$ & $  0$ & $ 0$ & $\neq 0$ & $\neq 0$ \\
\hline
\rule[-1mm]{0mm}{8mm}$C^4$& $C^2_c$ &$x(x^2+y^2)$ & $ 0$ & $ 0$ & $ 0$ & $ 0$ & $  0$ & $ 0$ & $\neq 0$ & $\neq 0$ \\
\hline
\rule[-1mm]{0mm}{8mm}$C^5$& $C^3_c$ &$x y^2$ & $ 0$ & $ 0$ & $ 0$ & $ 0$ & $  0$ & $ 0$ & $\neq 0$ & $ 0$ \\
\hline
\rule[-1mm]{0mm}{8mm}$C^6$& $C^4_c$ &$x^3$ & $ 0$ & $ 0$ & $ 0$ & $ 0$ & $  0$ & $ 0$ & $ 0$ & $ 0$ \\
\hline

\end{tabular}%
\end{center}
\caption{Classification of cubic curves with $I_{12}=0$}
\label{tab:clasinv}
\end{table}
\subsection{Classification over the reals}\label{sec:real}

For physical reasons what matters in applying these considerations to the $N=2$ Lagrangians and their corresponding non--linear BI limit is the classification over the reals. The classification over the reals reflects the $Sp(2n,R)$ symplectic structure of rigid special geometry \cite{S,SW,CDFvP}, with $GL(n,R)$ maximally embedded in $Sp(2n,R)$. Note that $PGL(n,R) \simeq SL(n,R)$. Now the different cases increase from 9 to 15, and in the following we are going to enumerate them.

To begin with, there are now two families with $I_{12} \neq 0$, since the sign matters. We shall refer to them as ``time--like'' $(\,\sigma < - \,\frac{1}{2}\,)$ and ``space--like'' $(\,\sigma > - \,\frac{1}{2}\,)$ orbits.  They all correspond to the representative polynomial given by the second canonical form \cite{JGS}:
\beq
U(x,y,z) \ = \ x^{\,3} \ + \ y^{\,3} \ + z^{\,3} \ + \ 6\, \sigma \, x\, y \, z \ , \qquad \left( \sigma \, \neq \, - \, \frac{1}{2} \right) \ ,
\eeq{125}
For these cases one has the following values of discriminant and invariants:
\beq
I_{12}\ =\ -\, 216\,(1\, +\, 8\, \sigma^3)^3 \ , \quad P_4\ = \ 24\,\sigma(-1+\sigma^3) \,  \quad Q_6\ = \ 6 \,(-1 + 20 \,\sigma^3 \, +\,  8\, \sigma^6) \ .
\eeq{126}

The remaining 13 cases refine the 8 cases of (A)--type, (B)--type and (C)--type degenerations, and the conditions on the invariant remain the same. Over the reals, however, the splitting goes as follows: there are 3 cases in (A), 4 cases in (B) and 6 cases in (C) {(see Table \ref{tab:clasinv}).}

In family (A) there are two options corresponding to a hyperbolic or an elliptic node, in family (B) the conic can be a circle or a hyperbola, and in case (C) a pair of lines can be real or made up of complex conjugates. The new real cases corresponding to $(P_4,Q_6) \neq (0,0)$ are distinguished by the sign of $Q_6$. There is one case for each of the (A), (B) and (C) families where this occurs, so that over the reals we have pairs of cases of type $(P_4,Q_6)$ and $(P_4,-Q_6)$. {Note that for $\sigma=-\frac{1}{2}$ the models with $I_{12} \neq 0$ degenerate into the case $C_2$.}

\section{Properties of the kinetic matrix}\label{sec:kinetic matrix}

In solving for the BI Lagrangians that correspond to the preceding classification, it is important to consider also the positivity properties of the $d_{AB}$ Hessian matrix of eq.~\eqref{0}, where $q^C = \Im X^C$.

As was the case for the $n=2$ constructions described in \cite{FPS}, in most of the examples the matrix $d_{AB}$ is not positive definite. An additional ingredient, the $C_{AB}$ matrix, is thus needed to have proper kinetic terms, which are determined in general by
\beq
\Im U_{AB} \ = \ C_{AB} \ + \ d_{ABC}\ \Im X^C \ .
\eeq{13}
However, there are some non--trivial cases where multiplet mixing occurs in which the matrix is positive definite, so that one can work with $C_{AB}=0$. This can be ascertained resorting to
Sylvester's criterion \cite{sylvester}, and thus analyzing the signs of all
diagonal minors of the Hessian matrices. In this fashion one can see that positivity holds in models with $I_{12} \neq 0$ for certain ranges of values for the $\sigma$ parameter and also in the model $B_1$ in Table \ref{tab:clasinv} (see Table \ref{tab:hessian}).

The models with $(P_4,Q_6) \neq (0,0)$ are generically more entangled. The triangle has $\partial^{\,2} I_{12} =0$, the line not tangent to the conic has $\partial I_{12} =0$, while the node has $\partial I_{12} \neq 0$.

For illustrative purposes, in the Appendix we give the solutions of the non--linear constraints \eqref{5} and \eqref{6} for one example with $(P_4,Q_6) \neq (0,0)$ of each family belonging to the classes (A), (B) and (C), as well as for the non--degenerate cases with $I_{12} \neq 0$. In class (A) the solution involves up to cubic radicals, in class (B) it arises from a biquadratic equation and in class (C) it simply involves ratios of simple radicals. Finally, the $I_{12} \neq 0$ cases result in a fourth--order equation. The degree of the equation that finally determines the solution is thus higher if the cubic is less degenerate, but in all cases the non--linear constraints have an algebraic solution.
\begin{table}[tbp]
\begin{center}
\begin{tabular}{|c||c|c|c|}
\hline
\rule[-1mm]{0mm}{6mm} $\rm Polynomial$ & $\rm Determinant$ & $\rm Minor2$ & $\rm Minor1$
\\ \hline\hline
\rule[-1mm]{0mm}{8mm}$x^3+y^3+z^3+6\, \sigma\, x \,y\, z$ & $x\, y\, z\,(1+2\,\sigma^3)$ & $ x y-z^2\, \sigma ^2$ & $x$\\
\rule[-1mm]{0mm}{8mm}$ $ & $-\left(x^3+y^3+z^3\right) \,
\sigma ^2 $ & $ $ & $ $\\
\hline
\rule[-1mm]{0mm}{8mm}$-x^3-x^2z+y^2z$ & $3 x y^2-x^2 z+y^2 z$ & $ -\frac{z}{9} \, (3 x+z)$ & $-x-\frac{z}{3}$\\
\hline
\rule[-1mm]{0mm}{8mm}$-x^3+x^2z+y^2z$ & $\frac{x y^2}{9}-\frac{1}{27} \left(x^2+y^2\right) z$ & $ \frac{z}{9} (-3 x+z)$ & $- x+\frac{z}{3}$\\
\hline
\rule[-1mm]{0mm}{8mm}$-x^3+y^2z$ & $\frac{x y^2}{9}$ & $-\frac{x z}{3}$ & $-x$\\
\hline
\rule[-1mm]{0mm}{8mm}$(x+y+z) \left(x^2+y^2+z^2\right)$ & $\frac{1}{27} (x+y+z)\left[x^2+y^2 \right.$ & $\frac{2}{9} \left(x^2+4 x y+y^2\right)$ & $x+\frac{y+z}{3}$\\
\rule[-1mm]{0mm}{6mm}$ $ & $ \left.+8 y z+z^2+8 x (y+z)\right]$ & $+\frac{1}{9}\left[4(x+y) z+z^2\right]$ & $ $\\
\hline
\rule[-1mm]{0mm}{8mm}$x \left(x^2+y^2-z^2\right)$ & $ -\frac{x}{27}  \left(3 x^2-y^2+z^2\right)$ & $ \frac{1}{9} \left(3 x^2-y^2\right)$ & $x$\\
\hline
\rule[-1mm]{0mm}{8mm}$z \left(x^2+y^2-z^2\right)$ & $ -\frac{z}{27} \left(x^2+y^2+3 z^2\right)$ & $ \frac{z^2}{9}$ & $\frac{z}{3}$\\
\hline
\rule[-1mm]{0mm}{8mm}$(x+z) \left(x^2+y^2-z^2\right)$ & $-\frac{4}{27}(x+z)^3$ & $ \frac{1}{9} \left[(x+z)(3 x+z)- y^2\right]$ & $ x+\frac{z}{3}$\\
\hline
\rule[-1mm]{0mm}{8mm}$6\,x\,y\,z$ & $2 \, x\, y\, z$ & $ -z^2$ & $0$\\
\hline
\rule[-1mm]{0mm}{8mm}$x \left(y^2+z^2\right)$ & $-\frac{x}{27} \left(y^2+z^2\right)$ & $-\frac{y^2}{9}$ & $0$\\
\hline
\rule[-1mm]{0mm}{6mm}$x\, y \, (x+y)$ & $0$ & $ \frac{1}{9} \left(-x^2-x y-y^2\right)$ & $\frac{y}{3}$\\
\hline
\rule[-1mm]{0mm}{8mm}$x \left(x^2+y^2\right)$ & $0$ & $ \frac{1}{9} \left(3 x^2-y^2\right)$ & $x$\\
\hline
\rule[-1mm]{0mm}{8mm}$x\,y^2$ & $0$ & $ -\frac{y^2}{9}$ & $0$\\
\hline
\rule[-1mm]{0mm}{8mm}$x^3$ & $0$ & $ 0$ & $x$\\
\hline

\end{tabular}%
\end{center}
\caption{Minors of the Hessian matrices}
\label{tab:hessian}
\end{table}

\section{Conclusions and outlook}\label{sec:conclusion}

In this investigation, which extends the results of ref.\cite{FPS}, we have constructed examples of $U(1)^3$ supersymmetric BI actions that are invariant under a second non--linearly realized supersymmetry. The complications met in solving the BI constraints (see eqs.~\eqref{4}--\eqref{6}) reflect the properties of the $d_{ABC}$ coefficients, which are related to particular choices of a projective cubic curve. Generically, the Hessian matrix $\partial_A \, \partial_B \, U$ of the cubic polynomial is not positive definite, which makes the introduction of a quadratic term depending on another matrix $C_{AB}$ necessary. Still, the non--linear BI action can be computed in this case as discussed in \cite{FPS}. For $n>3$ projective cubic $n-2$ varieties come into play, and a full classification of the resulting options is required.

Many related problems can be discussed in the context of non--linear theories of electro--magnetic fields. In particular, the corrections to the Coulomb force, which in standard BI can be computed exactly, or simply the bounds on the electric fields, which arise from reality requirements, deserve a further investigation. {However, we have gathered evidence that when the $d_{AB}$ matrix is positive definite the allowed electric fields are bounded in the absence of magnetic fields, in analogy with what happens for the standard Born--Infeld action. On the other hand, we have gathered evidence that some electric fields become unbounded when this condition does not hold.} Moreover, we expect that the $U(1)^n$ actions enjoy a self--duality property as in the $n=1$ case, but unlike in \cite{BMZ} we do not expect that the $U(1)^n$ duality extend to $U(n)$ as proposed in \cite{RT,BMZ,KT}. It would be clearly of interest to couple these BI systems to $N=2$ supergravity \cite{Ferrara:1976fu}, and also to clarify their relation to other brane systems \cite{Bergshoeff:2013pia}. ``Brane supersymmetry breaking'' \cite{bsb}, where all supersymmetries are non--linearly realized, was similarly connected to supergravity in \cite{dmpr}. The $N=4$ extensions of this class of models \cite{K,Bellucci:2000ft,Broedel,Bergshoeff:2013pia} can also be a fruitful area of investigation, with potentially important lessons for the ill--understood non--abelian generalization of the BI construction. Our experience with a similar problem, the coupling of the Volkov--Akulov \cite{VA} multiplet to supergravity, which was constructed in a similar language in \cite{ADFS}, makes us expect that it should be possible to couple these types of multiplets, which possess an $N=2$ non--linearly realized supersymmetry, to $N=2$ supergravity \cite{Freedman:2012zz}, and to explore the consequences of these interactions.

\subsection*{Acknowledgements} We are grateful to M.~Floratos and V.~S.~Varadarajan for enlightening discussions. S.~F. and A.~S. are supported by the ERC Advanced Investigator Grant n.~226455 (SUPERFIELDS). M.~P. is supported in part by NSF grant PHY-1316452. A.~Y. would like to thank the CERN Ph--Th Unit for the kind hospitality and the ERC Advanced Investigator Grant n. 226455 for support while at CERN. A.~S. is also supported in part by Scuola Normale Superiore and by INFN (I.S. Stefi). A.~S and R.~S. wish to thank the CERN Ph--Th Unit for the kind hospitality.
\begin{appendix}
\section{Explicit solutions of the constraints}\label{sec:appendix}
\end{appendix}
In this Appendix we consider the generalized BI Lagrangians for the non--degenerate family of eqs.~\eqref{125} and \eqref{126}, and for the less degenerate examples belonging to the three families (A), (B) and (C) in Table \ref{tab:clasinv}. The case of interest corresponding to the (A) family is $A_1$ in the first row ($I_{12}=0$),
and similarly for the (B) family it is $B_1$ in the fourth row ($\partial I_{12}=0$), while for the (C) family it is $C_1$ in the eight row ($\partial^{\,2} I_{12}=0$). Note that in all the ensuing analysis the weak--field limit provides unique root determinations with positive arguments. In that limit, in fact, the infinitesimal solutions $\delta \,H^A$, where
\beq
H^A \ = \ \frac{m^A}{2} \ + \ \delta \, H^A \ ,
\eeq{131}
are captured by the model--independent expressions
\beq
\delta\, H^A \ = \ \left( d_{AB}^{\,m} \right)^{\,-\,1}\, d_{BPQ} \ G^P \cdot G^Q \ ,
\eeq{132}
where
\beq
d_{AB}^{\,m} \ = \ d_{ABC}\ m^C \ ,
\eeq{133}
and $d_{AB}^{\,m}$ is always invertible in the three--field case (see Table \ref{tab:hessian}). An expression that is identical up to a sign gives the solution of eq.~\eqref{6} as
\beq
\Im F^A \ = \ - \ \left( d_{AB}^{\,m} \right)^{\,-\,1} \, d_{BPQ} \ G^P \cdot {\tilde{G}}^Q \ ,
\eeq{134}
since in the infinitesimal everything is linear.

Let us begin by considering the $I_{12}\neq 0$ case, where the representative polynomial and the corresponding invariants are given in eqs.~\eqref{125} and \eqref{126}.
Notice that $I_{12} \geq 0 $ for $\sigma \leq - \ \frac{1}{2}$, while it is negative in the remaining interval. The solutions are triality--symmetric and are determined by the zeroes of a fourth--order polynomial for $H^i$.
To begin with,
\begin{eqnarray}
\Im F^i &=&\frac{G^i\cdot\tilde{G^i}\,m^j\,m^k\ +\ A^i\,\sigma\ +\ B^i\,\sigma^2\ +\ C^i\,\sigma^3}{\left[(m^i)^3 \ + \ (m^j)^3 \ +\
(m^k)^3\right]\, \sigma^2 \ - \ m^i\, m^j\, m^k\, \left(1 + 2 \sigma^3\right)} \ ,
\end{eqnarray}
where $(i,j,k)=1,2,3$, $\left(i\neq j\neq k\right)$, and moreover
\begin{eqnarray}
A^i&=&2\, G^j\cdot\tilde{G^k}\,m^j\,m^k \ - \ G^j\cdot\tilde{G^j}\,(m^k)^2 \ - \ G^k\cdot\tilde{G^k}\,(m^j)^2 \ ,\nonumber\\
B^i&=&G^j\cdot\tilde{G^j}\, m^i\, m^j \ + \ G^k\cdot\tilde{G^k}\, m^i\, m^k \ - \ G^i\cdot\tilde{G^i}\, (m^i)^2 \nonumber\\
&&\hspace{2.5cm} - \ 2\, G^i\cdot\tilde{G^j}\, (m^j)^2 \ - \ 2\, G^i\cdot\tilde{G^k}\, (m^k)^2\ ,\nonumber\\
C^i&=&2\, m^i\, \left(-G^j\cdot\tilde{G^k}\, m^i \ + \ G^i\cdot\tilde{G^k}\, m^j \ +\ G^i\cdot\tilde{G^j}\, m^k\right)\ .
\end{eqnarray}
On the other hand
\beq
H^1\ =\ \sqrt{\frac{{A_{11}\, A_{22}}}{{A_{22}+2\, U\, \sigma  \left(A_{33}\, \sigma +\sqrt{A_{22}\, A_{33} \,U^2-2 A_{22}^2\, U\, \sigma +A_{33}^2\, \sigma ^2}\right)}}}\ ,
\eeq{H1i12}
where $U$ is a solution of the fourth--order equation
\begin{eqnarray}
&&U^4 \left(A_{11}^2\, -\, 4 \,A_{22}\, A_{33}\, \sigma ^2\right)\ +\ 4\, U^3 \, \sigma ^2 \left(A_{11}\, A_{33}\ +\ 2\, A_{22}^2 \, \sigma \right)\ -\ 2\, A_{11}\, A_{22} \,U^2 \left(1\ +\ 8 \, \sigma ^3\right) \nonumber\\&&+\ 4\, U \,\sigma ^2 \left(A_{22} \, A_{33}\, +\,2 \,A_{11}^2 \, \sigma \right)\ +\ A_{22}^2\ -\ 4\, A_{11} \, A_{33}\, \sigma ^2\ = \ 0
\end{eqnarray}
that is consistent with the weak--field limit.
Here
\begin{eqnarray}
A_{i\,i}\ = \ R^{i\,i}\ + \ 2\, \sigma R^{j\,k}\ ,
\end{eqnarray}
with $(i,j,k)=1,2,3$, $(i\neq j\neq k)$, and there are no implicit sums. The other components of $H^i$ can be obtained by a cyclic permutation of indices starting from the explicit form of eq.~\eqref{H1i12}. Two of these cases are simple: for $\sigma=0$, when the polynomial is diagonal, and for $\sigma=-\frac{1}{2}$, when $I_{12}=0$.

The other three examples are drawn from the degenerate cases in Table \ref{tab:clasinv}. The second example concerns $A_1$, and in this case the polynomial is
\beq
U(x,y,z) \ = \ -\ x^3 \ - \ z\left( x^2 \ - \ y^2 \right) \ .
\eeq{15}
The corresponding solution rests on a bi--cubic. To begin with, the explicit form of $\Im F^i$ reads
\begin{eqnarray}
\Im F^1&=&\frac{m^2}{M} \left(2\, G^2\cdot\tilde{G^3} \,m^1\ -\ 3 \, G^1\cdot\tilde{G^1} \, m^2\ - \ 2\, G^1\cdot\tilde{G^3} \, m^2\right) \nonumber\\
&&\hspace{5cm}+\frac{ m^1 m^3}{M}\left(G^1\cdot\tilde{G^1}-G^2\cdot\tilde{G^2}\right)\ ,\nonumber\\
\Im F^2&=&\frac{m^1}{M}\left (2 \, G^2\cdot\tilde{G^3} \, m^1\ -\ 2\, G^1\cdot\tilde{G^3} \, m^2\ -\ 3\, G^2\cdot\tilde{G^2} \, m^2\right) \nonumber\\
&&\hspace{5cm}+\ \frac{ m^2 m^3}{M} \left(G^1\cdot\tilde{G^1}\ -\ G^2\cdot\tilde{G^2}\right)\ , \\
\Im F^3&=&\frac{m^1}{M} \left(2 G^1\cdot\tilde{G^3} m^3+3 G^2\cdot\tilde{G^2} m^3-6 G^2\cdot\tilde{G^3} m^2\right) \nonumber\\
&&\hspace{3cm} +\ \frac{m^3}{M}\left(G^2\cdot\tilde{G^2}\, m^3\ -\ G^1 \cdot \tilde{G^1}\, m^3\ -\ 2\, G^2\cdot\tilde{G^3} \, m^2\right)\ , \nonumber
\end{eqnarray}
where
\beq
M \ = \ 3 \, m^1 \, (m^2)^2\ -\ (m^1)^2 \, m^3\ +\ (m^2)^2\, m^3 \ .
\eeq{155}
Moreover
\beq
H^2 \ = \ \frac{2\,R^{23}\,H^1}{3 \, R^{11} \ + \ 2 \, R^{13} \ - \ 3\, (H^1)^2}\ , \qquad H^3 \ = \ \frac{3 \, R^{11} \ + \ 2\,  R^{13} \ - \ 3\, (H^1)^2}{2\, H^1} \ ,
\eeq{16}
and
\begin{eqnarray}
H^1 \ = \ \frac{1}{3}\ \sqrt{C\ +\ 2\, \Re \left[ \, e^{\frac{2\pi i}{3}} \ \left( B \, - \, \sqrt{A}\,\right)^\frac{1}{3}\right]} \ ,
\end{eqnarray}
where
\begin{eqnarray}
A&=&-\, D^3\ +\ B^2\ ,\qquad
B\ =\ -\ (2\, R^{13}\ +\ 3\, R^{22})^3\ +\ 18\, C\, \left(R^{23}\right)^2 \ ,\nonumber\\
C&=&9\,R^{11}\ +\ 4\,R^{13}\ -\ 3\,R^{22}\ ,\qquad
D=\left(2\, R^{13}\ +\ 3\, R^{22}\right)^2\ +\ 12\, \left(R^{23}\right)^2\ .
\end{eqnarray}

The third example concerns $B_1$, and the associated polynomial reads
\beq
U \ = \ \left(x\,+\,y\,+\,z\right)\,\left(x^2\,+\,y^2\,+\,z^2\right) \ .
\eeq{18}
In this case the solutions for $H^i$ are determined essentially by a bi--quadratic equation. To begin with,
\begin{eqnarray}
\Im F^i&=&\frac{1}{\sum_n m^n \left[\sum_n (m^n)^2+8\left(\,m^i \,m^j\,+\,m^i\, m^k\,+\,m^j \,m^k\right)\right]}\ \biggl\{-G^i\cdot\tilde{G^i}\left[(m^i)^2  \right.\nonumber\\
&+& \left. 9 \, m^i \, m^j+ 6\, (m^j)^2 + 9 \, m^i \, m^k + 20\, m^j \, m^k + 6 \, (m^k)^2\right]+G^j\cdot\tilde{G^j}\left[3 (m^i)^2  \right.\nonumber\\
&+&\left. m^i \, m^j + 3\, m^i \, m^k - 4 (m^k)^2\right]+G^k\cdot\tilde{G^k}\left[3\, (m^i)^2 + m^i \, m^k + 3 \, m^i\, m^j - 4 (m^j)^2\right] \nonumber\\
&-& 2\, G^i\cdot\tilde{G^j}\left[ 3\, m^i \, m^j + (m^j)^2+ 2 \, m^i \, m^k + 6\,  m^j \, m^k + 3\, (m^k)^2\right] \nonumber\\
 &-& 2\, G^i\cdot\tilde{G^k}\left[ 3\, m^i \, m^k +  (m^k)^2+2\,  m^i \, m^j + 6\, m^j \, m^k + 3\, (m^j)^2\right] \nonumber\\
&+& 2\, G^j\cdot\tilde{G^k}\left[2 \, (m^i)^2 + 3\, m^i \, m^j + 3\, m^i \, m^k + 4\, m^j \, m^k\right]\biggr\}\ . \nonumber
\end{eqnarray}
Moreover, the explicit solution for $H^i$ takes the form
\begin{eqnarray}
H^i &=&  \frac{1}{6 \sqrt{3}} \ \sqrt{\frac{\alpha_{i} \ - \ \, {\beta}_{i}\, \sqrt{8 (A_i A_j+A_i A_k + A_j A_k )- 5 (A_i^2+A_j^2+A_k^2)}}{A_i^2+A_j^2+A_k^2-A_i A_j-A_i A_k- A_j A_k}} \ , \nonumber \\
\alpha_{i} &=& -7 (A_j^3+ A_k^3)+38 \, A_i^3
+3 \left[9 \,(A_j^2+A_k^2)-8 \, A_j\, A_k\right] A_i \nonumber \\
&&+9 \, (A_j+A_k) \left( A_j A_k- 4 \, A_i^2\right) \ , \nonumber \\
\beta_{i} &=& A_j^2+A_k^2+10\, A_i^2 +8\, A_k A_j -10\,
(A_j+A_k) A_i \ ,
\end{eqnarray}
where
\begin{eqnarray}
A_i\ = \ 3\, R^{i\,i} \ + \ R^{j\,j} \ +\ R^{k\,k} \ + \ 2\left( R^{i\,j} \ + \  R^{i\,k}\right)  \ ,
\end{eqnarray}
and $i,\,j,\,k$ run over $1,\,2,\,3$ and $i\neq j\neq k$.

Finally, the last non--trivial example concerns $C_1$ in Table \ref{tab:clasinv}, for which
\beq
U \ = \ 6\,x\,y\,z \ .
\eeq{19}
The explicit form of $\Im F^i$ is in this case
\begin{eqnarray}
\Im F^i \ = \ \frac{G^j\cdot\tilde{G^k}\, m^i\ -\ G^i\cdot\tilde{G^k}\, m^j\ -\ G^i\cdot\tilde{G^j}\,m^k}{
m^j\,m^k} \ ,
\end{eqnarray}
and moreover
\begin{eqnarray}
H^i\ =\ \frac{\sqrt{R^{i\,j}}\, \sqrt{R^{i\,k}}}{2\, \sqrt{R^{j\,k}}} \ ,
\end{eqnarray}
where the $R^{i\,j}$ are defined in eq.~(\ref{4}). The explicit form of $H^i$ thus reads
\begin{eqnarray}
H^i=\frac{\sqrt{-4\,\Im F^i\,\Im F^j+m^i\,m^j+4\, G^i\cdot G^j}\,\sqrt{-4\,\Im F^i\,\Im F^k+m^i\,m^k+4\, G^i\cdot G^k}}{2\,\sqrt{-4\,\Im F^j\,\Im F^k+m^j\,m^k+4\, G^j\cdot G^k}} \ ,
\end{eqnarray}
where $(i,\,j,\,k)= 1,\,2,\,3$ and $i\neq j\neq k$.

 In all other cases of Table \ref{tab:clasinv} the constraints can be solved in a similar way.

 {We have verified that the radial electric fields allowed for point sources are bounded in case 3, where a positive definite $d_{AB}$ matrix exists, and are unbounded in cases 2 and 4. The family of models with $I_{12} \neq 0$ clearly presents transition points between these two regimes, which depend on $\sigma$ and on the $m$--charges: close to $\sigma=0$ the allowed ranges are bounded, while for large positive or negative values of $\sigma$ they are unbounded.}


\end{document}